\documentclass[12pt]{article}
\usepackage{jheppub,graphicx}
\usepackage{amssymb}
\usepackage{amsmath, tensor}
\usepackage{microtype}
\usepackage{hyperref}

\newcommand{\half}[0]{\frac{1}{2}}
\newcommand{\cs}[3]{\Gamma^{#1}_{\,\,\,\, #2#3}}

\newcommand{\ld}[0]{\mathcal{L}}

\newcommand{\dd}[0]{\textrm{d}}
\newcommand{\defn}[0]{\equiv}

\newcommand{\qsubrm}[2]{{#1}_{\textrm{#2}}}

\newcommand{\hct}[0]{\mathcal{H}}
\renewcommand{\figurename}{Figure}

\def\be{\begin{align}}
\def\ee{\end{align}}
\def\bea#1\eea{\begin{align}#1\end{align}}
\def\bse{\begin{subequations}}
\def\ese{\end{subequations}}

\let\oldsqrt\sqrt
\def\sqrt{\mathpalette\DHLhksqrt}
\def\DHLhksqrt#1#2{%
\setbox0=\hbox{$#1\oldsqrt{#2\,}$}\dimen0=\ht0
\advance\dimen0-0.2\ht0
\setbox2=\hbox{\vrule height\ht0 depth -\dimen0}%
{\box0\lower0.4pt\box2}}

\newcommand{\CAMB}[0]{{\tt{CAMB}}}

\newcommand{\Mp}{\qsubrm{m}{P}^2}

\subheader{\today}
\title{Simple numerical implementation of general dark energy models}
\author[a]{Jolyon K. Bloomfield}
\author[b]{and Jonathan A. Pearson}
\affiliation[a]{MIT Kavli Institute for Astrophysics and Space Research, Massachusetts Institute of Technology, Cambridge, MA 02139, USA}
\emailAdd{jolyon@mit.edu}
\emailAdd{jonathan.pearson@durham.ac.uk}
\affiliation[b]{Centre for Particle Theory, Department of Mathematical Sciences, Durham University, South Road, Durham, DH1 3LE, U.K.}

\abstract{We present a formalism for the numerical implementation of general theories of dark energy, combining the computational simplicity of the equation of state for perturbations approach with the generality of the effective field theory approach. An effective fluid description is employed, based on a general action describing single-scalar field models. The formalism is developed from first principles, and constructed keeping the goal of a simple implementation into \CAMB\, in mind. Benefits of this approach include its straightforward implementation, the generality of the underlying theory, the fact that the evolved variables are physical quantities, and that model-independent phenomenological descriptions may be straightforwardly investigated. We hope this formulation will provide a powerful tool for the comparison of theoretical models of dark energy with observational data.}

\begin{document}
\maketitle
\section{Introduction}
Recent cosmological observations have concreted the notion of the dark sector: some unknown substance or new gravitational physics is currently dominating the gravitational dynamics of the universe \cite{Riess:1998cb, Perlmutter:1998np}. There are no shortage of theories which have been constructed in an attempt to describe these observations (see \cite{Copeland:2006wr, Clifton:2011jh} for reviews).

Given the recent deluge of data  \cite{Das:2011ak, vanEngelen:2012va, Bennett:2012zja, Ade:2013tyw, Planck:2013kta, Heymans:2012gg, Blake:2011en} and upcoming experiments \cite{Abbott:2005bi, Laureijs:2011mu, Amendola:2012ys}, it is of paramount importance that the understanding of dark sector theories is optimised for confrontation with observations. As such, it is becoming clear that some framework should be constructed which is capable of confronting entire classes of theories with data, and an   industry has begun to construct different flavours of such a formalism \cite{Weller:2003hw, Hu:2007pj, Skordis:2008vt, Baker:2011jy, Zuntz:2011aq, Baker:2012zs, Amendola:2012ky, Sawicki:2012re, Silvestri:2013ne, Baker:2013hia, Park:2010cw, Bloomfield:2011np, Mueller:2012kb}. Whatever the formalism is, it is important that it allows observations to be transcribed into well defined and meaningful statements about the allowed properties of the dark sector.

The aim of this paper is to present a leap forwards in the development of a model independent framework which can cover as much of the ``known theory space'' as possible. We do this by bringing together the effective field theory (EFT) approach to linear cosmological perturbations with the formalism for constructing the equations of state for perturbations. The former  approach \cite{Bloomfield:2012ff, Gubitosi:2012hu, Gleyzes:2013ooa, Bloomfield:2013efa} is well suited for incorporating complicated theories, whereas the latter \cite{Battye:2012eu, Pearson:2012kb, Battye:2013er, Battye:2013aaa} is optimized for comparing theories to data.

The ``holy grail'' would be to provide a parameterization that covered all of Horndeski's theory \cite{Horndeski:1974wa, Deffayet:2011gz, Kobayashi:2011nu}, since that is the most general single-scalar field theory with second order field equations. In this paper we provide the penultimate step in such an endeavour: our calculations and results are applicable to almost all of Horndeski's theory, although the formalism can be straightforwardly extended. Our emphasis is torn between theoretical generality and usability -- we provide a simple scheme for modifying numerical codes, such as \CAMB \, \cite{Lewis:1999bs}, which will allow observational spectra to be extracted from very complicated models with a minimum of effort. This is entirely due to the way that the equations of state for perturbations work: they modify the fluid equations with terms which are already evolved. We never explicitly have to evolve the scalar fields equation of motion.

We begin this paper by describing a framework in which non-minimally coupled theories may be decomposed into background terms, perturbations, and an effective fluid description in Section \ref{sec:darksector}. Next, the EFT approach is introduced in Section \ref{sec:EFT}. The equations of state for perturbations in the EFT model under consideration are developed in Section \ref{sec:EOS}, and the envisaged implementation of these results in a numerical code is laid out in Section \ref{sec:computation}. We conclude by discussing the benefits and drawbacks of this approach as well as future prospects in Section \ref{sec:discuss}. Appendix \ref{appendix:conventions} describes our conventions in detail, while Appendix \ref{appendix:explicit} provides complete explicit formulae for the equations of state.

\section{Dynamics of the dark sector}\label{sec:darksector}
We will be studying non-minimally coupled dark sector\footnote{By ``dark sector'', we mean dark energy and modified gravity; dark matter is assumed to be included in the matter Lagrangian.} theories. In this section we will lay down the basic framework, field equations, and associated conservation equations for dealing with a dark sector which is non-minimally coupled to gravity. We begin by laying everything out in tensorial notation with very general statements, before specializing to the context of cosmology. In subsequent sections we write down the relevant equations in component form.

\subsection{General case}
We organise the gravitational  action   into the general form
\begin{align}
S = \int \dd^4x\sqrt{-g}\, \bigg[ \frac{\Mp}{2} \Omega R - \qsubrm{\ld}{m} - \qsubrm{\ld}{d}\bigg],
\end{align}
where $\Omega$ is the coupling function, $\qsubrm{\ld}{m}$ is the matter Lagrangian,  and $\qsubrm{\ld}{d}$ is the dark Lagrangian, containing all dark sector fields. We chose to work in a conformal frame where there is no dark sector coupling to matter (for scalar field theories, this is the Jordan frame). This precludes the description of models that violate the weak equivalence principle, such as models with different couplings to matter and dark matter, within this framework. After varying the action with respect to the metric, the gravitational field equations are
\begin{align}
\Mp \Omega {G^{\mu}}_{\nu} = \big[{T^{\mu}}_{\nu} + {U^{\mu}}_{\nu}\big] - \Mp ({g^{\mu}}_{\nu}\square-\nabla^{\mu}\nabla_{\nu} )\Omega.
\label{eq:sec:grav-fld-eqns-coupled-prenet}
\end{align}
The matter and dark sector energy-momentum tensors (EMT) are defined respectively as
\begin{align}
T^{\mu\nu} \defn \frac{2}{\sqrt{-g}}\frac{\delta}{\delta g_{\mu\nu}}(\sqrt{-g}\qsubrm{\ld}{m}),\qquad U^{\mu\nu} \defn \frac{2}{\sqrt{-g}}\frac{\delta}{\delta g_{\mu\nu}}(\sqrt{-g}\qsubrm{\ld}{d}).
\end{align}
We now proceed by manipulating these general objects and defining various quantities so as to elucidate the effective fluid nature of the dark sector theory and how non-minimal couplings can be brought into the language of the fluid description.

We define a \textit{gross dark EMT} ${\mathfrak{U}^{\mu}}_{\nu}$ to include the dark sector EMT and the contribution from the derivatives of the coupling function. The definition is
\begin{align}
{\mathfrak{U}^{\mu}}_{\nu} \defn {U^{\mu}}_{\nu} - \Mp ({g^{\mu}}_{\nu}\square-\nabla^{\mu}\nabla_{\nu} )\Omega,
\end{align}
and the gravitational field equations (\ref{eq:sec:grav-fld-eqns-coupled-prenet}) become
\bea
\Mp \Omega {G^{\mu}}_{\nu}    = {T^{\mu}}_{\nu} + {\mathfrak{U}^{\mu}}_{\nu}.
\label{eq:sec:grav-fld-eqns-coupled}
\eea
The components of ${\mathfrak{U}^{\mu}}_{\nu}$ become the ``effective'' energy density and pressure of the dark sector. After taking into account the Bianchi identity $\nabla_{\mu}{G^{\mu}}_{\nu}=0$ and the conservation of energy for the matter sector $\nabla_{\mu}{T^{\mu}}_{\nu}=0$ (which follows from the choice of frame), it follows from (\ref{eq:sec:grav-fld-eqns-coupled}) that the  conservation equation in terms of  the gross dark EMT is given by
\bea
\label{eq:sec:cons-eqn-tensor-coupled}
\nabla_{\mu}{\mathfrak{U}^{\mu}}_{\nu} = \qsubrm{m}{P}^2 {G^{\mu}}_{\nu}\nabla_{\mu}\Omega.
\eea
The perturbed gravitational field equations are obtained from (\ref{eq:sec:grav-fld-eqns-coupled}) as
\bea
\label{eq:sec:pert-grav-eqns-tensor-coupled}
\Mp \Omega \delta {G^{\mu}}_{\nu}= \delta {T^{\mu}}_{\nu} + \delta {\mathfrak{U}^{\mu}}_{\nu} - \Mp {G^{\mu}}_{\nu}\delta \Omega.
\eea
As was the case on the background, it is insightful  to  combine the $\delta {\mathfrak{U}^{\mu}}_{\nu}$ and ${G^{\mu}}_{\nu}\delta \Omega$ contributions in the perturbed field equation (\ref{eq:sec:pert-grav-eqns-tensor-coupled}) into   the \textit{gross perturbed dark EMT},  ${\delta \mathcal{U}}{{}^{\mu}}_{\nu}$, which we define as
\bea
{\delta \mathcal{U}}{{}^{\mu}}_{\nu}\defn  \delta {\mathfrak{U}^{\mu}}_{\nu} -  \Mp {G^{\mu}}_{\nu}\delta \Omega.
\eea
This yields
\bea
\Mp \Omega \delta {G^{\mu}}_{\nu}= \delta {T^{\mu}}_{\nu} + \delta \tensor{\mathcal{U}}{^\mu_\nu}
\label{eq:sec:pert-grav-eqns-tensor-coupled-clean}
\eea
for the perturbed field equation.

The price we pay for this clean perturbed field equation is that the conservation equation for the gross perturbed dark EMT becomes somewhat complicated. Expanding the conservation equation \eqref{eq:sec:cons-eqn-tensor-coupled} to linear order in perturbations and making use of the perturbed field equation \eqref{eq:sec:pert-grav-eqns-tensor-coupled-clean} yields
\bea
\label{eq:sec:gross-pert-emt-tensor-coupled}
\nabla_{\mu}\delta{\mathcal{U}^{\mu}}_{\nu} = (\delta {T^{\mu}}_{\nu} + \delta{\mathcal{U}^{\mu}}_{\nu} ) \frac{1}{\Omega}\nabla_{\mu}\Omega  + 2{\mathfrak{U}^{\mu}}_{[\alpha}\delta\cs{\alpha}{\nu]}{\mu} ,
\eea
where $\delta\cs{\alpha}{\mu}{\nu} = g^{\alpha\beta}(\nabla_{(\mu}\delta g_{\nu)\beta} - \tfrac{1}{2}\nabla_{\beta}\delta g_{\mu\nu})$ is the perturbation to the connection symbols from metric perturbations. We see that the equations of motion for the gross perturbed dark EMT are sourced by not only themselves and metric perturbations, but also by ``normal'' matter perturbations. Note that the equations of motion for ``normal matter'' remain unchanged, however.

\subsection{Cosmology}
We wish to apply the above equations to a cosmological context\footnote{In order to avoid cluttering this section with our choice of conventions, we refer the reader to Appendix \ref{appendix:conventions}.}. On a background compatible with Friedmann-Robertson-Walker (FRW) symmetries (isotropy and homogeneity), the background gross dark EMT can be written as
\bea
\label{eq:sec:grossemt-defn}
{\mathfrak{U}^{\mu}}_{\nu} = (\rho + P) u^{\mu}u_{\nu} + P {\delta^\mu}_\nu,
\eea
where $\rho$ and $P$ are the energy density and pressure of the dark sector fluid, and $u^\mu$ is the fluid four-velocity (fluid variables without labels are fluid variables of the dark fluid). The components of the gross perturbed dark EMT $ {\delta \mathcal{U}}{{}^{\mu}}_{\nu}$ can be parameterized in a fluids language as
\bea
 {\delta \mathcal{U}}{{}^{\mu}}_{\nu} = (\delta \rho + \delta P) u^{\mu}u_{\nu} + \delta P {\delta^\mu}_\nu + (\rho + P) (v^\mu u_\nu + u^\mu v_\nu) + P \tensor{\Pi}{^\mu_\nu}.
\eea
We will refer to $\delta \rho$, $v^{\mu}$, $\delta P$ and ${\Pi^{\mu}}_{\nu}$ as the perturbed energy, velocity field, perturbed pressure and anisotropic stress of the dark sector fluid: these are the \textit{perturbed fluid variables of the dark sector}. If the dark sector contains only a single scalar field (without higher time derivatives), then there will only be two scalar components contained in these perturbed fluid variables, corresponding to the field and its time derivative. We later present relations between the perturbed fluid variables and the scalar field, as derived from an effective field theory action.

Using (\ref{eq:sec:grossemt-defn}), the background conservation equation (\ref{eq:sec:cons-eqn-tensor-coupled}) yields the coupled fluid equation
\bea
\dot{\rho} = - 3\hct (\rho+P) +3\dot{\Omega} \frac{\qsubrm{m}{P}^2}{a^2} (\hct^2 + k_0),
\eea
where $\hct\defn \dot{a}/a$ is the Hubble expansion, $k_0$ is the spatial curvature, and overdots represent derivatives with respect to conformal time. For scalar perturbations in the synchronous gauge in momentum space, the coupled perturbed fluid equations (\ref{eq:sec:gross-pert-emt-tensor-coupled}) become
\bse
\label{eq:sec:coupl_pert-fluid-eqns}
\bea
\label{eq:sec:coupl_pert-fluid-eqns-ddot}
\dot{\delta} = -(1+w)\bigg( \half \dot{h} + k^2\theta\bigg) - 3 \hct w\Upsilon +\frac{\dot{\Omega}}{\Omega} \bigg( \delta + \frac{\delta\qsubrm{\rho}{m}}{\rho}\bigg) ,
\eea
\bea
\label{eq:sec:coupl_pert-fluid-eqns-thetadot}
\dot{\theta} = - \hct(1-3w)\theta - \frac{\dot{w}}{1+w}\theta + \frac{w}{1+w}\left( \delta +\Upsilon - 2(\tfrac{1}{3}- \tfrac{  k_0}{k^2}) \Pi\right)   + \frac{\dot{\Omega}}{\Omega}\bigg(\theta+ \frac{\qsubrm{\rho}{m} + \qsubrm{P}{m}}{\rho+P}\qsubrm{\theta}{m} \bigg),
\eea
\ese
where $w = P/\rho$ is the dark sector ``equation of state'' and
\bea
\label{eq:sec:defn-wGamma}
w\Upsilon \equiv {\delta P}/{\rho} - w \delta.
\eea
Note that for dark energy with constant equation of state, $w \Upsilon$ is the entropy perturbation $w \Gamma$ of the dark sector fluid. Also note that these perturbed fluid equations are explicitly sourced by the perturbed matter fluid variables, due to the terms proportional to $\dot{\Omega}$. Finally, it is evident that this formulation cannot cope with models which cross the ``phantom divide'' of $w = -1$, and thus we require $w > -1$ for this formalism to make sense.

\section{Effective field theory model}\label{sec:EFT}
The perturbed fluid equations (\ref{eq:sec:coupl_pert-fluid-eqns}) are not closed, since we have not yet specified the form of the $w \Upsilon$ or anisotropic stress $\Pi$ perturbations. For a given single scalar dark energy model, it should be possible to relate both of these perturbations to the scalar field perturbation and its time derivative. In the following sections, we obtain the precise forms of these equations of state for perturbations from an effective field theory (EFT) action. In this section, we briefly summarise the EFT action we employ for coupled scalar field models.

Based on the \textit{Effective Field Theory of Inflation} formalism \cite{Cheung:2007st}, the \textit{Effective Field Theory of Dark Energy} \cite{Gubitosi:2012hu, Bloomfield:2012ff} has been used to describe perturbations in general single-scalar field models of dark energy. The action consists of three terms that determine the background evolution of the cosmology, and further operators that only contribute to the perturbative behavior of the scalar field. It has been demonstrated \cite{Bloomfield:2012ff, Gleyzes:2013ooa, Bloomfield:2013efa} that only three perturbative operators are required to completely describe perturbations in the most general single-scalar field theory with second order equations of motion, known as ``Horndeski's Theory'' \cite{Horndeski:1974wa} and recently rediscovered as ``Generalized Galileons'' \cite{Deffayet:2011gz, Kobayashi:2011nu}. Therefore, it is of interest to describe this action within the current formalism. However, the final term (with coefficient $\bar{M}_2^2$ in \cite{Bloomfield:2013efa}) significantly complicates the analysis we wish to perform, and thus we leave it for future work. The model without this term is sufficient to describe non-minimally coupled Kinetic Gravity Braiding models \cite{Kobayashi:2010cm, Deffayet:2010qz} (corresponding to the first two terms of Horndeski's theory, and a specialized combination of the second two).

The action in unitary gauge and conformal time\footnote{We are unaware of any work using the EFT of Inflation formalism that uses conformal time; the formalism is much more straightforward in physical time, but applies just as well in conformal time.} is given by
\begin{align}
  S = \int d^4x \sqrt{-g} \bigg\{ & \frac{\Mp}{2} \Omega(\tau) R
  + \Lambda(\tau)
  - c(\tau) a(\tau)^2 \delta g^{00}
  + \frac{M_2^4 (\tau)}{2} (a(\tau)^2 \delta g^{00})^2
\nonumber \\
  &
  - \frac{\bar{M}_1^3 (\tau)}{2} a(\tau)^2 \delta g^{00} \delta \tensor{K}{^i_i} - \qsubrm{\ld}{m} \bigg\}.
  \label{eq:sec:EFT-action}
\end{align}
The non-minimal coupling is implemented by the function $\Omega(\tau)$. The functions $\Lambda(\tau)$ and $c(\tau)$ along with $\Omega(\tau)$ are responsible for the dark sector's contribution to the evolution of the background, and the functions $M_2^4(\tau)$ and $\bar{M}_1^3(\tau)$ modify the evolution of linearized perturbations. The St\"uckelberg trick is used to restore gauge invariance in this action, introducing a scalar field $\pi$. The relevant diffeomorphism is $\tau \rightarrow \tau + \pi/a$, where we introduce the scalefactor for later convenience\footnote{The formalism we are pursuing here is invariant under field redefinitions, as it works with physical quantities rather than with fields. Thus, it it helpful to use the most straightforward definition.}.

The gravitational field equations are given by \eqref{eq:sec:grav-fld-eqns-coupled-prenet}. The $\Lambda(\tau)$ and $c(\tau)$ terms are the sole contributors to ${U^{\mu}}_{\nu}$ on the background. They are then combined with the derivatives of the coupling function $\Omega$ to yield the gross energy-momentum tensor ${\mathfrak{U}^{\mu}}_{\nu}$. Parameterizing the components of ${\mathfrak{U}^{\mu}}_{\nu}$ as (\ref{eq:sec:grossemt-defn}), the effective density and pressure of the dark sector fluid are \cite{Gubitosi:2012hu, Bloomfield:2012ff}
\bea
\rho = 2c - \Lambda -  \frac{\qsubrm{m}{P}^2}{a^2}3\hct{}\dot{\Omega}\,,\qquad P = \Lambda + \frac{\qsubrm{m}{P}^2}{a^2}(\ddot{\Omega} +  \hct\dot{\Omega})\,.
\eea
Solving these for $c$ and $\Lambda$ yield
\begin{align}
  c &=
  \frac{\rho + P}{2}
  - \frac{\Mp}{2a^2} \left(  \ddot{\Omega}-2\hct \dot{\Omega}
\right)
\,,&
  \Lambda &= P - \frac{\Mp}{a^2} \left(\ddot{\Omega}+\hct \dot{\Omega}   \right) \,.
\end{align}
The gravitational field equations (\ref{eq:sec:grav-fld-eqns-coupled}) and conservation equation (\ref{eq:sec:cons-eqn-tensor-coupled}) respectively become
\bse
\bea
\hct^2 &= \frac{a^2}{3\qsubrm{m}{P}^2\Omega}( \qsubrm{\rho}{m}+ \rho) - k_0\,,\\
6\dot{\hct} &= - \frac{a^2}{\qsubrm{m}{P}^2\Omega}(\qsubrm{\rho}{m}+\rho + 3\qsubrm{P}{m} + 3 P)\,,\\
\dot{\rho} &= - 3 \hct(\rho+P)+ \frac{3\qsubrm{m}{P}^2\dot{\Omega}}{a^2}(\hct^2+ k_0)\,.
\eea
\ese

The perturbed EMT for the action \eqref{eq:sec:EFT-action} can be computed directly, and have been checked against previously derived results transformed from physical to conformal time (see, e.g., the appendices of \cite{Bloomfield:2012ff}). We present the components of the perturbed field equations for different operators in the EFT, giving the gross perturbed dark EMT $\delta \tensor{\mathcal{U}}{^\mu_\nu}$ in momentum space (note the inclusion of mode functions, see Appendix \ref{appendix:conventions}).
\begin{itemize}
\item The perturbed EMT for operators $\Omega$, $c$ and $\Lambda$ is given by
\bse
\bea
  \delta \tensor{\mathcal{U}}{^0_0} &\supset \left[- \dot{\rho}  \frac{\pi}{a} - 2 c \frac{\dot{\pi}}{a} \right] Y
  + \frac{\qsubrm{m}{P}^2}{a^2} \dot{\Omega} \left[\bigg(k^2+3 (2 \hct^2 + k_0 - \dot{\hct})   \bigg)\frac{\pi}{a}+ 3 \hct \frac{\dot{\pi}}{a}+ \half{\dot{h}}\right] Y,
\\
  \delta \tensor{\mathcal{U}}{^0_i} &\supset k \left[\left(\rho + P\right) \frac{\pi}{a}+ \frac{\Mp}{a^2} \dot{\Omega} \frac{\dot{\pi}}{a}\right] Y_i,
\\
  \delta \tensor{\mathcal{U}}{^i_j} &\supset
  \left[\dot{P} \frac{\pi}{a}
  + \left(\rho + P\right) \frac{\dot{\pi}}{a}\right]{\delta^i}_j Y
  + \frac{\Mp}{a^2} \dot{\Omega} \left[\frac{1}{2} (\dot{h} + 6 \dot{\eta}) +   k^2 \frac{\pi}{a}\right] {Y^i}_j
\nonumber\\
  & \quad + \frac{\Mp}{a^2} \dot{\Omega} \left[\bigg( \left( 3 \hct^2 + k_0 \right)  + \frac{2}{3} k^2\bigg) \frac{\pi}{a}
    +\bigg( \hct + \frac{\ddot{\Omega}}{\dot{\Omega}}\bigg) \frac{\dot{\pi}}{a}
      + \frac{\ddot{\pi}}{a}
  + \frac{1}{3} \dot{h}
  \right] {\delta^i}_j Y.
\eea
\ese
\item The perturbed EMT for the operator $M_2^4$ is given by
\bse
\bea
  \delta \tensor{\mathcal{U}}{^0_0} &\supset - 4 M_2^4 \frac{\dot{\pi}}{a} Y,
\\
  \delta \tensor{\mathcal{U}}{^0_i} &\supset 0,
\\
  \delta \tensor{\mathcal{U}}{^i_j} &\supset 0.
\eea
\ese
\item The perturbed EMT for the operator $\bar{M}_1^3$ is given by
\bse
\begin{align}
  \delta \tensor{\mathcal{U}}{^0_0} &\supset \bar{M}_1^3 \left[\frac{\dot{h}}{2 a}+\frac{1}{a^2}\big(  3  {\hct^2  }{ } - 3  {\dot{\hct}  }{ }  +  {k^2  }{ }
\big)\pi  + 3 \frac{\hct}{a^2} \dot{\pi}
  \right] Y,
\\
  \delta \tensor{\mathcal{U}}{^0_i} &\supset \bar{M}_1^3 \frac{\dot{\pi}}{a^2} k Y_i,
\\
  \delta \tensor{\mathcal{U}}{^i_j} &\supset \frac{1}{a} (3 \hct + \partial_\tau) \left[
  \bar{M}_1^3 \frac{\dot{\pi}}{a} \right] {\delta^i}_j Y.
\end{align}
\ese
\end{itemize}

Now that we know the form of the dark EMT, we can read off the expressions for the perturbed fluid variables $\delta \rho$, $\delta P$, $\Pi$ and $\theta$.  This yields the following.
\bse
\label{eq:sec:perrt-fluid-vars-coup_kgb}
\begin{align}
\delta\rho &= \dot{\rho} \frac{1}{a}\pi
  + \frac{ 2 c }{a}\dot{\pi}
  - \frac{\Mp}{a^2} \dot{\Omega} \left[
 \frac{1}{a}\big(  k^2+3 [2 {\hct}^2 + k_0 - \dot{\hct}] \big)\pi  +   {\hct} \frac{3}{a}\dot{\pi}
  + \frac{1}{2}\dot{h}
  \right]\nonumber\\&{}\qquad
  + 4 M_2^4 \frac{\dot{\pi}}{a}
  - \bar{M}_1^3 \left[ \frac{1}{a^2}\big(  3  {\hct^2  }{ } - 3  {\dot{\hct}  }{ }  +  {k^2  }{ }
\big)\pi
  + 3 \frac{{\hct}}{a^2} \dot{\pi}+\frac{1}{2 a}\dot{h} \right]
\end{align}
\begin{align}
(\rho+P)\theta &=      \left(\rho + P\right) \frac{1}{a}\pi
  + \frac{\Mp}{a^2} \dot{\Omega} \frac{1}{a}\dot{\pi}
  + \bar{M}_1^3 \frac{1}{a^2} \dot{\pi}
\end{align}
\begin{align}\delta P &=   \dot{P} \frac{1}{a}\pi
  + \left(\rho + P\right) \frac{1}{a}\dot{\pi}
  + \frac{1}{a} (3 {\hct} + \partial_\tau) \left[
  \bar{M}_1^3 \frac{\dot{\pi}}{a} \right]
  \nonumber\\
  &\qquad
  + \frac{\Mp}{a^2} \dot{\Omega} \bigg[
\frac{1}{a}\big(    3 {\hct}^2 + k_0
  + \tfrac{2}{3} k^2 \big)\pi
  + \frac{1}{a}\bigg( {\hct}   + \frac{\ddot{\Omega}}{\dot{\Omega}}   \bigg) \dot{\pi}
  + \frac{1}{a}\ddot{\pi}
  + \frac{1}{3} \dot{h}
  \bigg]
\end{align}
\begin{align}P\Pi &= \frac{\Mp}{a^2} \dot{\Omega} \frac{1}{2} (\dot{h} + 6 \dot{\eta}) + \frac{\Mp}{a^2} \dot{\Omega} k^2 \frac{1}{a}\pi
\end{align}
\ese
These formulae prescribe the explicit way in which the field variables and coefficients from the EFT action combine to modify each component of the perturbed gravitational field equations.

\section{The equations of state for perturbations}\label{sec:EOS}
We now move on to providing the equations of state for perturbations for these coupled scalar field theories. The key thing to realize is that the perturbed fluid equations (\ref{eq:sec:coupl_pert-fluid-eqns}) are not closed until both $w \Upsilon$ and the anisotropic stress $\Pi$ are specified as functions of fluid and metric components which already have evolution equations ($\delta$, $\theta$, $\dot{h}$ and $\eta$). In this section we show how this can be done. The first step in the calculation is to write the dark perturbed fluid variables (\ref{eq:sec:perrt-fluid-vars-coup_kgb}) as
\bea
\label{eq:sec:dpfvars-5}
\left( \begin{array}{c}\delta + A_{14}\dot{h} \\ \theta \\ \delta P - A_{34} \dot{h}  \\ \Pi - A_{44}(\dot{h} + 6 \dot{\eta})\end{array} \right) = \left( \begin{array}{ccc} A_{11} & A_{12} & 0 \\ A_{21}& A_{22} & 0 \\ A_{31} & A_{32} & A_{33} \\ A_{41} & 0 &0\end{array}\right)\left( \begin{array}{c}\pi \\ \dot{\pi} \\ \ddot{\pi}\end{array} \right).
\eea
The $[\qsubrm{A}{IJ}]$ are the components of what we call the ``activation matrix'', and can all be calculated in terms of the coefficients in the effective action (they may also depend on wavenumber); they are given explicitly in Appendix \ref{sec:mapp-ea-am}. For models that are more general than what we are considering here, there may be more non-zero components, as well as more metric perturbations appearing in the column vector on the left-hand-side. However, note that $A_{13}$ and $A_{23}$ will always be zero for single-scalar field models with second order equations of motion. In the model under consideration, it transpires that only $A_{11}$, $A_{31}$ and $A_{41}$ have wavenumber dependence.

The next step is to solve for $\pi$ and $\dot{\pi}$ in terms of $\delta$ and $\theta$ (and any metric perturbations). We begin with the following subexpression of the activation matrix (\ref{eq:sec:dpfvars-5}):
\begin{align}
\label{eq:sec:sub-exp-actmat-1}
  \left(
  \begin{array}{c}
    \delta + A_{14} \dot{h} \\
    \theta
  \end{array}
  \right)
  =
  \left(
  \begin{array}{cc}
    A_{11} & A_{12} \\
    A_{21} & A_{22}
  \end{array}
  \right)
  \left(
  \begin{array}{c}
    \pi \\
    \dot{\pi}
  \end{array}
  \right).
\end{align}
This can be easily inverted to give
\bse
\bea
\label{eq:pi-solved}
  \pi &= \frac{A_{22} (\delta + A_{14} \dot{h}) - A_{12} \theta}{\mathcal{D}} \,,
\\
\label{eq:pidot-solved}
  \dot{\pi} &= \frac{A_{11} \theta - A_{21} (\delta + A_{14} \dot{h})}{\mathcal{D}} \,,
\eea
\ese
where we defined
\bea
\label{eq:sec:denom_D-defn}
\mathcal{D} \defn A_{11}A_{22} - A_{12}A_{21}.
\eea
The denominator here should not   vanish. We give the explicit form of the denominator as a function of terms in the effective action  in (\ref{denom_p_pd}).

We now need an expression for $\ddot{\pi}$. Unfortunately, it's not as straightforward as just taking the time derivative of the expression for $\dot{\pi}$, as this would introduce $\dot{\delta}$ and $\dot{\theta}$ into the expression for $\delta P$. Using the equations of motion to eliminate $\dot{\delta}$ and $\dot{\theta}$ would in turn introduce $\delta P$ (because of the dependence of $w \Upsilon$ on $\delta P$), resulting in a circular definition. Instead, note from \eqref{eq:sec:coupl_pert-fluid-eqns} that the combination $\dot{\delta} + 3 \hct(1+w) \dot{\theta}$ is independent of $w \Upsilon$, and thus $\delta P$. Thus, we wish to obtain two expressions for $\ddot{\pi}$, and take the appropriate linear combination of them to precisely give this combination of the fluid derivatives.
Taking the time derivative of (\ref{eq:sec:sub-exp-actmat-1}) provides
\begin{align}
\label{eq:sec:eos_pre_2}
  \left(
  \begin{array}{c}
    \dot{\delta} + \dot{A}_{14} \dot{h} + A_{14} \ddot{h} \\
    \dot{\theta}
  \end{array}
  \right)
  =
  \left(
  \begin{array}{cc}
    \dot{A}_{11} & \dot{A}_{12} \\
    \dot{A}_{21} & \dot{A}_{22}
  \end{array}
  \right)
  \left(
  \begin{array}{c}
    \pi \\
    \dot{\pi}
  \end{array}
  \right)
  +
  \left(
  \begin{array}{cc}
    A_{11} & A_{12} \\
    A_{21} & A_{22}
  \end{array}
  \right)
  \left(
  \begin{array}{c}
    \dot{\pi} \\
    \ddot{\pi}
  \end{array}
  \right) \,.
\end{align}
In order to obtain the desired combination,  (\ref{eq:sec:eos_pre_2}) should be contracted with the row vector $[1, 3 \hct (1 + w)]$. The resulting equation can be solved for $\ddot{\pi}$, and yields
\begin{align}
\label{eq:pidotdot-solved}
  \ddot{\pi} = \frac{1}{ \mathcal{E} } \bigg( &  \dot{\delta} + 3 \hct (1 + w) \dot{\theta} +\dot{A}_{14}\dot{h} +  A_{14} \ddot{h} -  \mathcal{F}\pi - \mathcal{G}\dot{\pi}
  \bigg)
\end{align}
where we defined the expression appearing in the denominator as
\bea	
\label{eq:sec:denom_E-defn}
 \mathcal{E} \defn A_{12} + 3 \hct (1 + w) A_{22}
\eea
and the expressions appearing in the numerator as
\bse
\label{eq:sec:num_FG-defn}
\bea
\label{eq:sec:num_F-defn}
\mathcal{F} &\defn \dot{A}_{11} + 3 \hct (1 + w) \dot{A}_{21}\,,\\
\label{eq:sec:num_G-defn}
 \mathcal{G} &\defn A_{11} + \dot{A}_{12} + 3 \hct (1 + w) (A_{21} +  \dot{A}_{22})\,.
\eea
\ese

We now have expressions for $\pi, \dot{\pi}$ and $\ddot{\pi}$ in terms of perturbed fluid and metric variables: these are (\ref{eq:pi-solved}), (\ref{eq:pidot-solved}) and (\ref{eq:pidotdot-solved}) respectively. Inserting these expressions into the relevant slots in (\ref{eq:sec:dpfvars-5}) yields the following (schematic) expressions.
\bse
\label{eq:main-eos-result}
\bea
\label{eq:main-eos-result-Dp}
\delta P &= \mathcal{J}_1 \delta +   \mathcal{J}_2 \theta + \mathcal{J}_3 \dot{h} + \mathcal{J}_4 \ddot{h} + \mathcal{J}_5 [\dot{\delta} + 3 \hct (1 + w) \dot{\theta}]\\
\label{eq:main-eos-result-Dpi}
\Pi &= \mathcal{K}_1 \delta + \mathcal{K}_2 \theta + \mathcal{K}_3 \dot{h} + \mathcal{K}_4 \dot{\eta}
\eea
\ese
These are our \textit{equations of state for perturbations}. The coefficients $\mathcal{J}_i$ and $\mathcal{K}_i$ are defined in terms of the $\qsubrm{A}{IJ}$ in Appendix \ref{sec:appenx-comp-exp}. In the following section, we combine these equations with the Einstein equations to demonstrate how these equations of state can be numerically evolved.

\section{Computational steps}\label{sec:computation}
We envisage the implementation of these equations in software such as \CAMB\,  \cite{Lewis:1999bs}. In this section, we detail how we think such an implementation might work.

In the scalar sector, the metric perturbation $\eta$, the matter sector $\qsubrm{\delta}{m}$ and $\qsubrm{\theta}{m}$, and the dark sector $\delta$ and $\theta$ are known. The matter sector behavior should be unchanged. We assume that all quantities that depend only on the background are known, or can be computed as necessary.

The perturbed Einstein equations (in synchronous gauge, as used in \CAMB) are as follows.
\bse
\begin{align}
  \frac{\Mp \Omega}{a^2} \left[ - \hct \dot{h} + 2 k^2 \eta - 6 k_0 \eta\right]
&=
  - \delta \qsubrm{\rho}{m}
  - \delta \rho
\\
  \frac{\Mp \Omega}{a^2} \left[2 k^2 \dot{\eta} - k_0 (\dot{h} + 6 \dot{\eta}) \right]
&=
  (\qsubrm{\rho}{m} + \qsubrm{P}{m}) k^2 \qsubrm{\theta}{m}
  + (\rho + P) k^2 \theta
\\
  \frac{\Mp \Omega}{a^2}
  \left[
  \frac{2}{3} k^2 \eta
  - \frac{1}{3} \ddot{h}
  - \frac{2}{3} \hct \dot{h}
  - 2 \eta k_0
  \right]
&=
  \delta \qsubrm{P}{m} + \delta P \label{eq:einstein:pressure}
\\
  \frac{\Mp}{a^2} \Omega \left[
  k^2 \eta
  - \hct (\dot{h} + 6 \dot{\eta})
  - \frac{1}{2} (\ddot{h} + 6 \ddot{\eta})
  \right]
&=
  \qsubrm{P}{m} \qsubrm{\Pi}{m} + P \Pi \label{eq:einstein:anisotropic}
\end{align}
\ese
Recall that all fluid variables without labels are those for the dark sector.
From the first of these equations, we can compute $\dot{h}$. The second equation then supplies $\dot{\eta}$.

In order to proceed, we should now evaluate the denominators $\mathcal{D}$ and $\mathcal{E}$, and the numerators $\mathcal{F}$ and $\mathcal{G}$.
Expressions for these are all given in Appendix \ref{app:frequent}. Using these quantities, we can evaluate $\pi$ and $\dot{\pi}$ from \eqref{eq:pi-solved} and \eqref{eq:pidot-solved}.

Next, we can compute the anisotropic shear stress $\Pi$. From the results of the previous section, it is given by (\ref{eq:main-eos-result-Dpi}), although it may be easier to implement in terms of $\pi$ using the expression from \eqref{eq:def-aij-effaction-pi}. Among other things, this provides the driving term in   \eqref{eq:einstein:anisotropic} above.

We can now compute
\begin{align}
  \dot{\delta} + 3 \hct (1 + w) \dot{\theta}
\end{align}
which is also given in Appendix \ref{app:frequent}. The explicit expression for $\delta P$ can be written in terms of these various quantities as
\begin{align}
  \delta P &= A_{34} \dot{h} + A_{31} \pi + A_{32} \dot{\pi}
  + \frac{A_{33}}{ \mathcal{E} } \big(  \dot{\delta} + 3 \hct (1 + w) \dot{\theta} +\dot{A}_{14}\dot{h} +  A_{14} \ddot{h} -  \mathcal{F}\pi - \mathcal{G}\dot{\pi}
  \big) .
\end{align}
Unfortunately, we do not yet know $\ddot{h}$. For the moment, it makes sense to compute the quantity
\begin{align}
  \Delta P\defn \delta P - \frac{A_{33}}{\mathcal{E}} A_{14} \ddot{h} \,.
\end{align}
We now use \eqref{eq:einstein:pressure}; some simple rearrangements provide
\begin{align}
  \ddot{h}
  =
  \left(\frac{2}{3} k^2 \eta
  - \frac{2}{3} \hct \dot{h}
  - 2 \eta k_0
  - \frac{a^2}{\Mp \Omega} (\delta \qsubrm{P}{m} + \Delta P)\right) \Bigg/
  \left(\frac{a^2}{\Mp \Omega} \frac{A_{33} A_{14}}{\mathcal{E}} + \frac{1}{3} \right)
  \,.
\end{align}
It can be shown that the quantity in the denominator here is always positive. From here, it is now straightforward to construct $\delta P$.

We are now in a position to calculate
\begin{align}
  w \Upsilon = \frac{\delta P}{\rho} - w \delta\,,
\end{align}
which is the final quantity that needed to be computed. Now we have everything   needed to evolve the dark sector fluid equations \eqref{eq:sec:coupl_pert-fluid-eqns}.

\section{Discussion}\label{sec:discuss}
The formalism presented here suggests a way in which the dynamics of perturbations in reasonably complicated scalar field theories can be implemented in software such as \CAMB. There are a number of benefits to this approach, as well as some drawbacks.

The primary benefit of this approach is that the formalism is both general and self-contained. The action \eqref{eq:sec:EFT-action} that we started from is very general in that it contains most of the generality included in Horndeski's action, and can furthermore be extended to the full theory. At the same time, the implementation of the equations of motion needs to modify only a small handful of equations in \CAMB, as it already includes a simple quintessence implementation with constant equation of state. The computation of all the necessary coefficients for the formalism described here is slightly tedious, but not complicated.

A second major benefit to this formalism is that the quantities that are evolved in this formalism are physical quantities that appear in the EMT for dark energy. Thus, the evolution of physical quantities is straightforward to extract from the formalism, as is the meaning of the objects being evolved. Previous efforts to directly implement the equations of motion for the $\pi$ field ran into difficulties associated with the meaning of the $\pi$ field as a time displacement for the background scalar field, which is not a particularly meaningful physical quantity. Our formalism is also independent of redefinitions of the $\pi$ field, and so further removes any ambiguity associated with scalar fields.

An interesting point that may be considered a benefit or a drawback depending on your point of view is that the background evolution of the model must be precomputed in order to construct the various functions of time appearing in the action, as well as their derivatives. While this requires solving the background evolution of specific models, it also allows general phenomenological models to be described, for example, by choosing various functional forms for the coefficients. As a side note, we refer the reader to \cite{Bloomfield:2013efa}, which relates all of the coefficients appearing in the EFT action to functions appearing in Horndeski's action.

A further interesting point is that we have developed this formalism in conformal time using synchronous gauge. This is primarily motivated by the use of these choices in \CAMB, although other choices could of course be made. It may be interesting to investigate what the equations of state look like in a gauge-invariant formulation, in order to make more sense of the structures involved.

One of the drawbacks of this approach is that we cannot consider models which cross (or even touch) the ``phantom divide'' of $w = -1$. Doing so causes a number of denominators to become zero, which is clearly problematic. Although it is possible for models not to display instabilities in this regime, the description in terms of an effective fluid breaks down for such cases.

The other drawback to this approach is that we are ignoring the effects of kinetic mixing. In a number of models, including non-minimally coupled models and kinetic gravity braiding models \cite{Kobayashi:2010cm, Deffayet:2010qz}, the scalar degrees of freedom in the metric are kinetically mixed with the scalar field. To isolate the physical scalar degree of freedom requires the diagonalisation of the kinetic matrix, which allows the speed of sound to be calculated appropriately. The formalism described here simply lumps all of the kinetic mixing terms into the dark sector effective density and velocity field perturbations. As such, it is difficult to extract the speed of sound from this formalism, and one needs to remember that the scalar metric perturbations are not completely nondynamical. Nonetheless, the evolution equations presented here are exactly equivalent to the original equations of motion. We also note that the equations of motion in terms of the original scalar field can be used to obtain such properties in a complementary manner to this approach.

We believe that this formalism will be useful for comparing models to observational data. In particular, we hope to develop a tool by which theorists can rapidly compute  CMB and weak lensing spectra (and other perturbative effects) of models, without having to understand the intricacies of modifying \CAMB. The next step in this direction will be to expand these results to describe the full Horndeski theory, before diving into the computational realm.

\subsection*{Acknowledgments}
\vspace{-0.1in}
We thank Richard Battye, Rachel Bean, Eanna Flanagan, Adam Moss, and Ira Wasserman for useful conversations. JB was supported in part by NSF grants PHY-1068541 and PHY-0968820. JAP  is supported by the STFC Consolidated Grant ST/J000426/1.

\appendix
\section{Conventions}
\label{appendix:conventions}
\subsection{Metric}
We write the background FRW metric using conformal time as
\begin{align}
  ds^2 = a(\tau)^2 \left[-d\tau^2 + \tilde{g}_{ij} dx^i dx^j \right]\,,
\end{align}
where $\tilde{g}_{ij}$ is a three-dimensional spatial metric with no time dependence, such that the Riemann tensor of this spatial metric is given by
\begin{align}
  \tilde{R}_{ijkl} = k_0 \left(\tilde{g}_{ik} \tilde{g}_{jl} - \tilde{g}_{il} \tilde{g}_{jk} \right)\,,
\end{align}
where $k_0$ is the curvature constant (we reserve $k$ for wavenumber, and use $K$ for the trace of the extrinsic curvature tensor). We use overdots to denote derivatives with respect to conformal time.

We consider perturbations in synchronous gauge. The metric with perturbations is written as
\begin{align}
  ds^2 = a(\tau)^2 \left[-d\tau^2 + (\tilde{g}_{ij} + h_{ij}) dx^i dx^j\right].
\end{align}
The metric perturbation $h_{ij}$ decomposes into two scalar components as
\begin{align}
  h_{ij} = \frac{h}{3} \tilde{g}_{ij} + \left( \tilde{\nabla}_i \tilde{\nabla}_j - \frac{\tilde{g}_{ij}}{3} \tilde{g}^{kl} \tilde{\nabla}_k \tilde{\nabla}_l \right) \tilde{\eta}
\end{align}
where $\tilde{\nabla}$ is the covariant derivative associated with the spatial metric. Evidently, $\tilde{g}^{ij} h_{ij} = h$.

\subsection{Momentum space}
We follow the conventions of Kodama and Sasaki \cite{Kodama:1985bj}. For each wavevector $\vec{k}$, define $Y_{\vec{k}}(x^i)$ to be a solution of the equation
\begin{align}
  \tilde{\nabla}^2 Y_{\vec{k}} = - k^2 Y_{\vec{k}}.
\end{align}
From now onwards, we suppress the $\vec{k}$ dependence of $Y$. Taking derivatives of $Y$, we define vector and tensor mode functions as
\begin{align}
  Y_i &= - k^{-1} \tilde{\nabla}_i Y\,, &
  Y_{ij} &= k^{-2} \tilde{\nabla}_i \tilde{\nabla}_j Y + \frac{1}{3} \tilde{g}_{ij} Y.
\end{align}
We raise and lower indices on $Y_i$ and $Y_{ij}$ with the spatial metric only. This ensures that $Y$, $Y_i$, and $Y_{ij}$ are independent of time, no matter the position of their indices.

In synchronous gauge, the metric perturbation decomposes as
\begin{align}
  h_{ij} (\vec{x}, t) = \int d^3 k \left[ \frac{h(\vec{k}, t)}{3} Y \tilde{g}_{ij} + k^2 \tilde{\eta}(\vec{k}, t) Y_{ij} \right].
\end{align}
Following Ma and Bertschinger \cite{Ma:1995ey}, we define $\eta$ by
\begin{align}
  \tilde{\eta}(\vec{k}, t) &= - \frac{1}{k^2} [ h(\vec{k}, t) + 6 \eta (\vec{k}, t)].
\end{align}
Using this definition, the metric perturbation in synchronous gauge becomes
\begin{align}
  h_{ij} = \int d^3 k \left[ \frac{h}{3} Y \tilde{g}_{ij} - (h + 6 \eta) Y_{ij} \right].
\end{align}
In the limit of no spatial curvature, the mode function $Y = \exp(i \vec{k} \cdot \vec{x})$ yields Ma and Bertschinger's Eq. (4).

\subsection{Energy-momentum tensors}
On an FRW background, the background EMT for any sector can be written as
\begin{align}
  \tensor{T}{^\mu_\nu} = P {\delta^\mu}_\nu + (P + \rho) u^\mu u_\nu
\end{align}
where $u^\mu$ is the velocity vector of the fluid, $u^\mu = (1 / a, 0, 0, 0)$, and $P$ and $\rho$ are the background pressure and energy density. Quantities that belong to the matter sector are given a subscript ${}_{\mathrm{m}}$, while quantities associated with the dark sector have no subscripts.

Again following Kodama and Sasaki, we describe scalar perturbations of the EMT in momentum space using the following general decomposition (we use this decomposition for both the matter and dark sectors).
\begin{align}
  \delta \tensor{T}{^0_0} &= - \delta \rho Y
\\
  \delta \tensor{T}{^0_i} &= \rho (1 + w) v Y_i
\\
  \delta \tensor{T}{^i_j} &= \delta P Y{ \delta^i}_j + P \Pi \tensor{Y}{^i_j}
\end{align}
Note that indices on the modefunctions have been raised only with the spatial metric. We can convert $v$ into the $\theta$ variable of Ma and Bertschinger by $\theta = k v$. Similarly, Ma and Bertshinger's $\sigma$ variable for the anisotropic shear stress is related to $\Pi$ by $\sigma = 2 \Pi w /3 (1 + w)$. We find it more convenient to work with $\theta = v / k = \theta_{\mathrm{MB}} / k^2$, which we use throughout the rest of this document. Note that we use $\delta = \delta \rho / \rho$ as the fractional density perturbation.

\section{Explicit forms of coefficients}\label{appendix:explicit}
In this appendix, we present the explicit expressions for coefficients appearing in our formulae that we kept largely hidden in the main body of the paper.

\subsection{Mapping from the effective action to activation matrix}
\label{sec:mapp-ea-am}
We begin by presenting the components of the activation matrix \eqref{eq:sec:dpfvars-5} in terms of the quantities derived from the effective action \eqref{eq:sec:perrt-fluid-vars-coup_kgb}. Note that the inverse powers of the scalefactor are largely due to using conformal time. Common elements to a number of these quantities are
\begin{align}
  \mathcal{B} &= \Mp \frac{\dot{\Omega}}{a} + \bar{M}_1^3 \,,
\\
  \mathcal{C} &= \Mp \frac{\dot{\Omega}}{a^2} \,.
\end{align}

The perturbed density is given by
\bse
\label{eq:def-aij-effaction}
\begin{align}
  \delta  &= A_{11} \pi + A_{12} \dot{\pi} - A_{14} \dot{h},
\end{align}
where the activation matrix components are given by
\begin{align}
  A_{11} &= - \frac{1}{\rho} \left[3 \frac{\hct}{a} (\rho+P) + \frac{\mathcal{B}}{a^2} \left( k^2 + 3 ({\hct}^2 - \dot{\hct}) \right)  \right],
\\
  A_{12} &= \frac{1}{a \rho} \left[2 c + 4 M_2^4 - \frac{3 \hct \mathcal{B}}{a} \right],
\\
  A_{14} &= \frac{\mathcal{B}}{2 a \rho}.
\end{align}
\ese
The velocity divergence field is given by
\bse
\label{eq:def-aij-effaction-theta}
\begin{align}
  \theta &= A_{21} \pi + A_{22} \dot{\pi},
\end{align}
where the relevant activation matrix components are
\begin{align}
  A_{21} &= \frac{1}{a},
\\
  A_{22} &= \frac{\mathcal{B}}{a^2 (\rho + P)}.
\end{align}
\ese
The perturbed pressure is
\bse
\label{eq:def-aij-effaction-dp}
\begin{align}
  \delta P &= A_{31} \pi + A_{32} \dot{\pi} + A_{33} \ddot{\pi} + A_{34} \dot{h},
\end{align}
with
\begin{align}
  A_{31} &= \frac{\dot{P}}{a} + \frac{\mathcal{C}}{a} \left[ 3 {\hct}^2 + k_0 + \frac{2}{3} k^2 \right],
\\
  A_{32} &= \frac{\rho + 2P - \Lambda}{a}
  +\frac{1}{a^2}\bigg[  {(\bar{M}_1^3)^\cdot}{ }+{2 {\hct} \bar{M}_1^3}{ }
    \bigg],
\\
  A_{33} &= \frac{\mathcal{B}}{a^2},
\\
  A_{34} &= \frac{\mathcal{C}}{3}.
\end{align}
\ese
Here we denoted $(\bar{M}_1^3)^\cdot \equiv \partial_\tau \bar{M}_1^3$. Finally, the anisotropic stress is
\bse
\label{eq:def-aij-effaction-pi}
\begin{align}
  \Pi &= A_{41} \pi + A_{44} (\dot{h} + 6 \dot{\eta}),
\end{align}
with
\begin{align}
  A_{41} &= \frac{\mathcal{C}}{P a} k^2,
\\
  A_{44} &= \frac{\mathcal{C}}{2 P}.
\end{align}
\ese

\subsection{Frequently used expressions} \label{app:frequent}
We now show how the explicit expressions for the $\qsubrm{A}{IJ}$, which we gave as functions of terms in the effective action in (\ref{eq:def-aij-effaction} -- \ref{eq:def-aij-effaction-pi}), combine to yield some of the commonly appearing expressions in the coefficients in $\delta P$ and $\Pi$.

We begin with explicit expressions that appear in denominators, $\mathcal{D}$ and $\mathcal{E}$, defined in (\ref{eq:sec:denom_D-defn}) and (\ref{eq:sec:denom_E-defn}) respectively. We find
\begin{align}
\mathcal{E} = \frac{2}{a \rho} ( c + 2 M_2^4 ),
\end{align}
\begin{align}
\label{denom_p_pd}
 - a^2 \rho\mathcal{D} &=
  a \rho {\mathcal E}
  + \frac{\mathcal{B}^2}{a^2 \rho(1+w)} \left( k^2 + 3 ({\hct}^2 - \dot{\hct}) \right).
\end{align}
These expressions appear in the denominators of $\pi$, $\dot{\pi}$ and $\ddot{\pi}$, and as such, must be nonvanishing. In particular, since it is known from analysing the EFT construction that the stability of the theory requires $c + 2 M_2^4 > 0$, we expect both of these expressions to be positive.

We now move on to the explicit expressions for the terms $\mathcal{F}$ and $\mathcal{G}$, defined in (\ref{eq:sec:num_FG-defn}), as well as $\dot{A}_{14}$, which appear in numerators. These are by far the most complicated expressions we have to deal with. Beginning with $\mathcal{G}$, we break it up into
\bse
\begin{align}
\label{eq:sec:g-exp-1}
\mathcal{G}\supset  A_{11} + 3 \hct (1 + w) A_{21} = \frac{\mathcal{B}}{a^2 \rho} \left( 3 (\dot{\hct} - \hct^2) - k^2 \right),
\end{align}
\begin{align}
\label{eq:sec:g-exp-2}
\mathcal{G}\supset  \dot{A}_{12} + 3 \hct (1 + w) \dot{A}_{22} &=
  \frac{2}{a \rho} \left( \dot{c} + 2 (M_2^4)^\cdot
  - \left(\frac{\dot{\rho}}{\rho} + \hct \right) ( c + 2 M_2^4 ) \right)
\nonumber\\&\qquad
  - \frac{3 \mathcal{B}}{a^2 \rho} \left(\dot{\hct} + \frac{\hct \dot{w}}{1 + w}\right),
\end{align}
\ese
so that $\mathcal{G}$ is the sum of (\ref{eq:sec:g-exp-1}) and (\ref{eq:sec:g-exp-2}). We also have
\begin{align}
\mathcal{F}\defn  \dot{A}_{11} + 3 \hct (1 + w) \dot{A}_{21}
  &=
  - \frac{\mathcal{B}}{a^2 \rho} \left(\frac{\dot{\rho}}{\rho} + 2 \hct \right) \left( 3 (\dot{\hct} - \hct^2) - k^2 \right)
\nonumber\\&\qquad
  + \frac{1}{a^2 \rho} \left( \frac{\Mp \ddot{\Omega}}{a} - \frac{\Mp \hct \dot{\Omega}}{a} + (\bar{M}_1^3)^\cdot \right) \left( 3 (\dot{\hct} - \hct^2) - k^2 \right)
\nonumber\\&\qquad
  + \frac{\mathcal{B}}{a^2 \rho} \left( 3 \ddot{\hct} - 6 \hct \dot{\hct} \right)
  - \frac{3}{a} [\hct \dot{w} + \dot{\hct} (1 + w)] \,,
\end{align}
and
\begin{align}
  \dot{A}_{14} &=
  \frac{1}{2 a \rho} \left[
  \frac{\Mp}{a} \left(\ddot{\Omega}+\hct \dot{\Omega}   \right) + (\bar{M}_1^3)^\cdot
  + 2 \hct \bar{M}_1^3
  + \frac{\mathcal{B}}{\rho} \left(
  3 \hct w\rho
  - \frac{\dot{\Omega}}{\Omega}\left[ \rho +  {\qsubrm{\rho}{m}}{ } \right]
  \right)
  \right].
\end{align}

Finally, by combining (\ref{eq:sec:coupl_pert-fluid-eqns-ddot}) and (\ref{eq:sec:coupl_pert-fluid-eqns-thetadot}) we obtain
\begin{align}
\label{eq:app:fluiddotexpr}
\dot{\delta} + 3 \hct(1+w) \dot{\theta} &=
  -(1+w)\bigg( \half \dot{h} + k^2\theta\bigg)
  - 3 \hct^2 (1+w) (1-3w) \theta - 3 \hct \dot{w} \theta
\nonumber\\&\qquad
  + 3 \hct w \left( \delta - \frac{2}{3} \Pi + \frac{2 k_0}{k^2} \Pi\right)
  + \frac{\dot{\Omega}}{\Omega} \bigg( \delta + \frac{\delta\qsubrm{\rho}{m}}{\rho}\bigg)
\nonumber\\&\qquad
  + 3 \hct (1+w)\frac{\dot{\Omega}}{\Omega}\bigg(\theta+ \frac{\qsubrm{\rho}{m} + \qsubrm{P}{m}}{\rho(1 + w)}\qsubrm{\theta}{m} \bigg) \,.
\end{align}

\subsection{Complete expressions} \label{app:complete}
\label{sec:appenx-comp-exp}
For completeness' sake, here we include the full expressions for $\Pi$ and $\delta P$. They can be written schematically as
\bea
\delta P &= \mathcal{J}_1 \delta +   \mathcal{J}_2 \theta + \mathcal{J}_3 \dot{h} + \mathcal{J}_4 \ddot{h} + \mathcal{J}_5 [\dot{\delta} + 3 \hct (1 + w) \dot{\theta}]\\
\Pi &= \mathcal{K}_1 \delta + \mathcal{K}_2 \theta + \mathcal{K}_3 \dot{h} + \mathcal{K}_4 \dot{\eta}.
\eea
This is before rearranging the space-space trace equation to obtain $\ddot{h}$. Note that the term with coefficient $\mathcal{J}_5$ can be expressed in terms of $\delta$, $\theta$ and $\Pi$, but it seems unnecessary to insert the explicit expression \eqref{eq:app:fluiddotexpr} that displays no dependence on $\qsubrm{A}{IJ}$ apart from that implicitly included in $\Pi$.

We find that the coefficients $\mathcal{J}_i$ are given in terms of the $\qsubrm{A}{IJ}$ as
\bse
\label{eq:sec:defn_Bi_eos_deltaP}
\begin{align}
  \mathcal{J}_1 &= \left(A_{31} - \frac{A_{33} \mathcal{F}}{\mathcal{E}}\right) \frac{A_{22}}{\mathcal{D}} - \left(A_{32} - \frac{A_{33} \mathcal{G}}{\mathcal{E}} \right) \frac{A_{21}}{\mathcal{D}}
\,,\\
  \mathcal{J}_2 &= \left(A_{32} - \frac{A_{33} \mathcal{G}}{\mathcal{E}} \right) \frac{A_{11}}{\mathcal{D}} - \left(A_{31} - \frac{A_{33} \mathcal{F}}{\mathcal{E}}\right) \frac{A_{12}}{\mathcal{D}}
\,,\\
  \mathcal{J}_3 &= A_{34} + \frac{A_{33} \dot{A}_{14}}{\mathcal{E}} + \left(A_{31} - \frac{A_{33} \mathcal{F}}{\mathcal{E}}\right) \frac{A_{22} A_{14}}{\mathcal{D}}
  - \left(A_{32} - \frac{A_{33} \mathcal{G}}{\mathcal{E}} \right) \frac{A_{21} A_{14}}{\mathcal{D}}
\,,\\
  \mathcal{J}_4 &= \frac{A_{33} A_{14}}{\mathcal{E}}
\,,\\
  \mathcal{J}_5 &= \frac{A_{33}}{\mathcal{E}}
\,.
\end{align}
\ese

Similarly, the coefficients $\mathcal{K}_i$ are given by
\bse
\label{eq:Sec:c-Pi-defns}
\bea
\mathcal{K}_1 &= \frac{A_{41}A_{22}}{\mathcal{D}}\,,
\\
\mathcal{K}_2 &= - \frac{A_{41}A_{12}}{\mathcal{D}}\,,
\\
\mathcal{K}_3 &= \frac{A_{41}A_{22}A_{14}}{\mathcal{D}} + A_{44}\,,
\\
\mathcal{K}_4 &= 6A_{44}\,.
\eea
\ese

\bibliographystyle{JHEP}
\bibliography{refs}
\end{document}